\documentclass[sigconf]{acmart}

\AtBeginDocument{%
  }

\copyrightyear{2026}
\acmYear{2026}
\setcopyright{cc}
\setcctype{by}
\acmConference[RecSys '26]{20th ACM Conference on Recommender Systems}{September 27-October 02, 2026}{Minneapolis, MN, USA}
\acmBooktitle{20th ACM Conference on Recommender Systems (RecSys '26), September 27-October 02, 2026, Minneapolis, MN, USA}
\acmDOI{10.1145/3773078.3831780}
\acmISBN{979-8-4007-2284-4/2026/09}

\usepackage{enumitem}
\usepackage{makecell}
\usepackage{xcolor}
\usepackage{hyperref} 
\usepackage{cleveref}  
\usepackage{algorithm}
\usepackage{algpseudocode}
\usepackage{amsmath}
\usepackage{graphicx} 
\usepackage{booktabs}
\usepackage{multirow}
\usepackage{pifont}
\newcommand{\cmark}{\ding{51}}
 
\makeatletter
\algrenewcommand\algorithmicrequire{\textbf{Input:}}
\algrenewcommand\algorithmicensure{\textbf{Output:}}
\makeatother

\begin{document}

\title{Topology-Aware Tokenization for Generative Recommendation}

\author{Yaokun Liu}
\affiliation{%
  \institution{University of Illinois Urbana-Champaign}
  \city{Champaign}
  \country{United States}
}
\email{yaokunl2@illinois.edu}

\author{Yifan Liu}
\affiliation{%
  \institution{University of Illinois Urbana-Champaign}
  \city{Champaign}
  \country{United States}
}
\email{yifan40@illinois.edu}

\author{Zhenrui Yue}
\affiliation{%
  \institution{University of Illinois Urbana-Champaign}
  \city{Champaign}
  \country{United States}
}
\email{zhenrui3@illinois.edu}

\author{Gyuseok Lee}
\affiliation{%
  \institution{University of Illinois Urbana-Champaign}
  \city{Champaign}
  \country{United States}
}
\email{gyuseok2@illinois.edu}

\author{Zelin Li}
\affiliation{%
  \institution{University of Illinois Urbana-Champaign}
  \city{Champaign}
  \country{United States}
}
\email{zelin3@illinois.edu}

\author{Ruichen Yao}
\affiliation{%
  \institution{University of Illinois Urbana-Champaign}
  \city{Champaign}
  \country{United States}
}
\email{ryao8@illinois.edu}

\author{Dong Wang}
\affiliation{%
  \institution{University of Illinois Urbana-Champaign}
  \city{Champaign}
  \country{United States}
}
\email{dwang24@illinois.edu}

\renewcommand{\shortauthors}{Yaokun Liu et al.}

\begin{abstract}
Generative recommendation reformulates sequential recommendation as an autoregressive generation task, yet a critical issue in this paradigm remains overlooked: topology distortion in item tokenization.
In particular, we observe that the intrinsic adjacency relationships of items in the pretrained semantic embedding space are significantly disrupted after quantization. This topology distortion misleads the model’s perception of item similarity, ultimately bottlenecking the accuracy of generative recommendations.
To address this issue, we propose \textbf{Topo}logy-Aware \textbf{Tok}enization (\textbf{TopoTok}), an item tokenization framework that preserves item relational structure throughout the quantization hierarchy. 
Different from the prior monolithic supervision in tokenization, TopoTok introduces a multi-level distillation scheme to progressively recover the topology from coarse to fine granularity: 1) Inter-Group Distillation to capture global cluster-wise relations; 2) Intra-Group Distillation to refine local structures within semantic clusters; and 3) Inter-Item Distillation to enforce fine-grained alignment at the individual item level. 
Extensive experiments on three benchmark datasets demonstrate that TopoTok effectively alleviates topology distortion and consistently outperforms state-of-the-art tokenizers, achieving significant performance gains of up to 9.42\% in Recall@5. 
\end{abstract}



\maketitle

\section{INTRODUCTION}\label{sec:intro}

\begin{figure}[t]
  \centering
  \includegraphics[width=\linewidth]{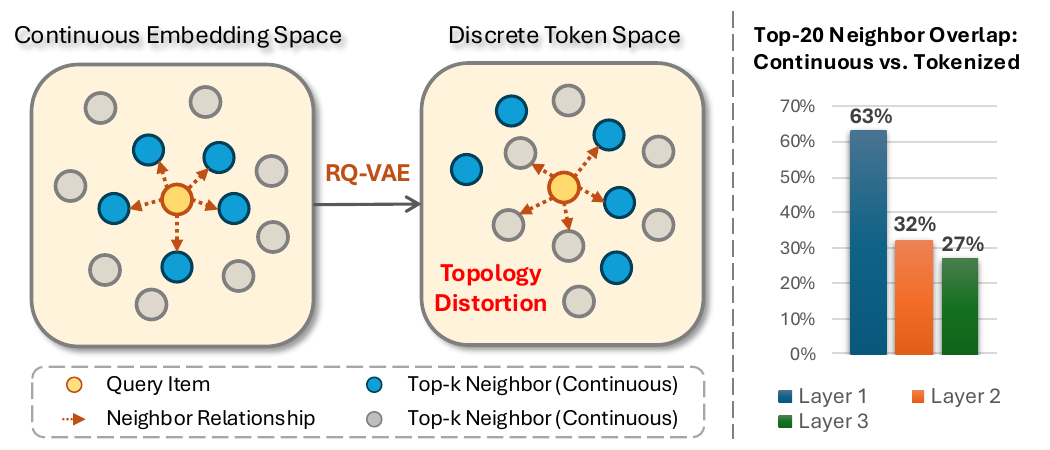}
  \caption{
  Illustration of topology distortion in item tokenization. 
  Left: Item adjacency relationships in the continuous embedding space are compromised after RQ-VAE quantization. 
  Right: Top-20 neighbor overlap between continuous and discrete spaces decreases with quantization depth.
  }
  \Description{A framework diagram illustrating topology distortion in semantic item tokenization. The left panel shows that items close together in the continuous semantic embedding space become separated and mixed with non-neighbors after RQ-VAE quantization into the discrete token space. The right panel displays a line chart showing that the Top-20 neighbor overlap between the continuous and tokenized spaces drops sharply from 63 percent at Layer 1, to 32 percent at Layer 2, and down to 27 percent at Layer 3.}
  \label{fig:intro}
  \vspace{-0.5cm}
\end{figure}

The advancement of Large Language Models (LLMs) has catalyzed a paradigm shift in sequential recommendation, moving from traditional embedding-based retrieval~\cite{dnn4rec, sasrec, bert4rec, s3rec} to generative recommendation~\cite{tiger, lcrec, cllm4rec, liu2025learning}. By reframing sequential recommendation as an autoregressive generation task, LLMs can directly generate target item identifiers, effectively bypassing the scalability bottlenecks and retrieval overhead of traditional Approximate Nearest Neighbor (ANN) search in high-dimensional spaces~\cite{tiger}. 
{Within the generative recommendation framework, item tokenization plays a critical role by transforming continuous item semantics into discrete identifiers that serve as generation targets, and its quality directly governs the model's ability to perceive item relationships and the ultimate accuracy of generative predictions.}

Early item tokenization methods fall into two categories: (1) \textit{Pseudo ID-based methods}, which assign each item a unique identifier via techniques such as random indexing~\cite{p5, index, cllm4rec}, offering high efficiency but lacking semantic information~\cite{tan2024towards}; and (2) \textit{text-based methods}, which replace item IDs with textual content (e.g., titles or descriptions) and formulate interactions as natural language prompts~\cite{tallrec, zeroshot, index}, capturing rich semantics but incurring high computational cost and potentially producing invalid outputs~\cite{index}.

{To balance the semantic richness of text-based representations and the efficiency of numerical IDs, recent work has focused on \textbf{semantic ID-based item tokenization}~\cite{tiger}, which maps continuous item embeddings into discrete token sequences, preserving semantic structure for autoregressive generation.}
{Within this paradigm, the Residual Quantized Variational Autoencoder (RQ-VAE)~\cite{rqvae} has become a dominant tokenizer due to its end-to-end learnability and 
ability to capture complex semantics. RQ-VAE employs a multi-layer codebook that hierarchically quantizes residual representations, encoding items from coarse to fine granularity. For example, in Figure~\ref{fig:pipeline}, a digital piano is tokenized into $<6, 3, 5>$, corresponding to instruments, keyboards, and digital pianos.}

{However, we identify a critical \textbf{topology distortion} problem that emerges from semantic ID-based tokenization workflows, {where we define topology as the relational structure among item representations induced by pairwise similarities (e.g., neighborhood ranking in the embedding space).}} 
{While continuous item embeddings provide rich semantic information for recommendation by capturing the intrinsic relational structure among items, this essential knowledge is progressively disrupted during the residual quantization process.} As illustrated in Figure~\ref{fig:intro}, our empirical analysis reveals that for the top-20 neighbor relationships defined in the continuous embedding space, only 63\% are preserved at layer 1, and this proportion further drops to 27\% at layer 3 during RQ-VAE tokenization. Such distorted relational structures introduce "topological noise" into the item identifiers, where semantically non-neighboring items are mistakenly treated as similar, fundamentally compromising the LLM's perception of item relationships and degrading autoregressive prediction accuracy.

{
In this paper, we adopt \textit{topology distillation} to transfer relational structure from the semantic embedding space to the item token space, directly addressing topology distortion by preserving item relationships.
However, existing relational knowledge distillation methods are not directly applicable in this setting, as they often assume homogeneous continuous embedding spaces, whereas topology preservation under discrete quantization introduces fundamentally different constraints~\cite{kang2021topology}.
Specifically, effective distillation within the residual quantization hierarchy faces two key challenges:}
{(\textit{C1}) \textbf{Layer-wise supervision.} The hierarchical architecture of RQ-VAE requires supervision to be applied at multiple layers; otherwise, a monolithic signal leads to supervision entanglement, where early layers are undertrained and deeper layers overfit to compensate for accumulated distortion.
(\textit{C2}) \textbf{Granularity alignment.} Since RQ-VAE encodes semantics from coarse to fine, the distillation granularity must align with each layer’s semantic role; enforcing fine-grained constraints uniformly can hinder early layers from capturing broader semantic structures.}

{
To this end, we propose \textbf{Topo}logy-Aware \textbf{Tok}enization (\textbf{TopoTok}), 
a hierarchical tokenization framework with layer-aligned topology supervision that preserves topological fidelity throughout RQ-VAE-based item tokenization. Specifically, we propose a layer-wise supervision scheme that decomposes topology distillation into multi-level objectives that can be integrated into the RQ-VAE hierarchy (\textit{C1}) with aligned coarse-to-fine semantic granularity (\textit{C2}). Specifically, TopoTok utilizes three levels of topology distillation: 
(1) \textbf{Inter-Group Distillation} captures coarse-grained topology at early layers by aligning similarities between group-level representations;  
(2) \textbf{Intra-Group Distillation} refines local structure at intermediate layers by preserving relations within semantic groups; 
(3) \textbf{Inter-Item Distillation} enforces fine-grained alignment at the final layer via item-level similarity matching.
}
This modular design allows TopoTok to be flexibly adapted to RQ-VAE architectures with an arbitrary number of quantization layers by mapping these distillation levels to the corresponding stages of semantic refinement.
Collectively, the hierarchical supervision of TopoTok empowers each quantization layer to distill topological priors at a layer-specific granularity, effectively mitigating topology distortion across the codebook and ensuring that discrete token identifiers remain structurally consistent with the original semantic manifold. Extensive experiments on three benchmarks demonstrate that TopoTok improves topology preservation and outperforms existing item tokenization methods for generative recommendation.

\section{RELATED WORK}

\subsection{Generative Recommendation}

In recent years, generative recommendation has emerged as a paradigm that formulates sequential recommendation as an autoregressive generation task. In contrast to traditional embedding-based retrieval methods, which typically rely on a two-tower model to compute ranking scores followed by efficient MIPS or ANN search for top-k retrieval~\cite{ge2013optimized,houle2014rank,jegou2010product,muja2014scalable}, generative recommenders leverage the context understanding capability of LLMs to generate the identifier tokens of the next item directly, enabling more flexible modeling and better handling of challenges such as cold start~\cite{genrecsys_review}. 
Early effort P5~\cite{p5} fine-tunes a pretrained language model T5~\cite{t5} to handle multiple recommendation tasks in a single model. Another seminal work is TIGER~\cite{tiger}, which proposes representing each item by a sequence of discrete semantic codes derived from item side information and then using a pretrained T5 to predict the next item’s codes. Subsequent research has enhanced the generative recommendation by integrating more signals and improving training strategies. For example, EAGER~\cite{eager} employs a two-stream generation framework to incorporate both user behavior history and item content semantics. ED$^2$~\cite{yin2025unleash} leverages an end-to-end unified framework that integrates semantic and collaborative filtering indexes using a multi-grained token regulator and task-specific instruction tuning. 
Existing generative recommenders often overlook topological supervision at the item tokenization stage, resulting in item token representations with distorted item relations. Our work addresses this gap by integrating topological distillation supervision into the item tokenization process.

\subsection{Item Tokenization}
A key challenge in generative recommendation is designing item tokenizations that LLMs can both interpret and generate effectively. Existing approaches fall into three main categories: pseudo ID-based~\cite{chu2023leveraging,p5,index,wang2024enhanced}, text-based~\cite{bao2025bi,dai2023uncovering,li2023generative,liao2023llara,zhang2025recommendation,zhang2021language,liao2024llara}, and semantic ID-based~\cite{tiger,lcrec}. 
{Pseudo ID-based methods, such as P5~\cite{p5}, assign unique tokens via techniques like random indexing, which are efficient but fail to capture intrinsic item-relatedness. Text-based methods~\cite{tallrec} utilize item metadata to reformulate recommendations as instruction-following tasks. 
While expressive, text-based methods incur high computational costs and are prone to hallucination~\cite{index}. }
{To bridge these gaps, semantic ID-based methods quantize item embeddings into discrete codes, preserving semantics in a structured form. For example,  TIGER~\cite{tiger} pioneered the use of RQ-VAE~\cite{rqvae} as a backbone, leveraging its hierarchical codebook to encode items from coarse to fine granularity. Since its inception, RQ-VAE has emerged as the predominant framework for semantic ID tokenization. LETTER~\cite{letter} integrates collaborative signals and diversity regularizations into tokenization. CoST~\cite{cost} introduces contrastive loss to maintain neighborhood relationships but supervises only the final output, leading to entangled constraints across quantization layers.  ETEGRec~\cite{etegrec} exploits the end-to-end learnability of RQ-VAE to jointly optimize the tokenizer and the recommender.
In industrial contexts, OneRec~\cite{onerec} alternatively uses RQ-KMeans to offer lightweight quantization. However, such non-parametric methods often struggle to capture the complex dependencies inherent in item semantics and lack the differentiability required for gradient-based joint optimization. In contrast, our work harnesses the differentiability of RQ-VAE to align topology distillation with the hierarchical semantic progression of residual quantization. This ensures that topological priors are systematically integrated into each stage of the codebook hierarchy in a coarse-to-fine manner.}

\section{METHODOLOGY}

\subsection{Preliminary}
\subsubsection{Problem Formulation}
We formulate the generative recommendation task under the sequential recommendation scenario. Given the set of items $\mathcal{I}$ and a user interaction sequence $\mathbf{S}^{u} = [i_{1}, i_{2}, \dots, i_{t-1}]\in \mathcal{I}$, the task is to predict the next item $i_{t} \in \mathcal{I}$.
Generative recommendation addresses this task through two key steps: item tokenization and autoregressive generation. Item tokenization maps each item $i \in \mathcal{I}$ into a token sequence $\mathbf{c}_i = [c_{i,1}, c_{i,2}, \dots, c_{i,L}] \in \mathcal{C}$, where $L$ is the sequence length and $\mathcal{C}$ is predefined token set. The user interaction sequence $\mathbf{S}^{u}$ is thereby transformed into a tokenized sequence $\mathbf{X}^{u} = [\mathbf{c}_{i_1}, \mathbf{c}_{i_2}, \dots, \mathbf{c}_{i_{t-1}}]$.
Given $\mathbf{X}^{u}$, the model autoregressively generates the token sequence $\mathbf{c}_{i_t}$ of the next item $i_t$ by factorizing the conditional probability as:
\begin{equation}
p(\mathbf{c}_{i_t}|\mathbf{X}^{u}) = \prod_{l=1}^{L} p(c_{i_t,l}|\mathbf{X}^{u}, c_{i_t,1}, \dots, c_{i_t,{l-1}}).
\end{equation}

\subsubsection{RQ-VAE for Semantic ID Tokenization}

Semantic ID-based item tokenization maps continuous item embeddings into discrete numerical tokens while preserving their semantics. RQ-VAE is widely adopted as the backbone model for hierarchical semantic encoding through multi-layer residual quantization. 

Given an item $i$, we first obtain its embedding $\mathbf{s}_i \in \mathbb{R}^{d_s}$ with pre-trained encoders such as SASRec~\cite{sasrec} or LLaMA~\cite{llama}, which captures collaborative or textual signals of items. 
The embedding $\mathbf{s}_i$ is then projected into a latent space by an MLP encoder:
\begin{equation}
\mathbf{z}_i = \mathrm{Encoder}(\mathbf{s}_i), \quad \mathbf{z}_i \in \mathbb{R}^{d_c}.
\end{equation}
The latent representation $\mathbf{z}_i$ is then quantized by a sequence of $L$ hierarchical codebooks $\{ \mathcal{C}_1, \dots, \mathcal{C}_L \}$, where each codebook $\mathcal{C}_l$ contains $N$ learnable code embeddings $\{\mathbf{e}_{l,j} \in \mathbb{R}^{d_c} \}_{j=1}^N$. 
The quantization at layer $l$ is performed through recursive residual mapping:
\begin{equation}\label{eq:residual_quant}
    c_{l} = \arg\min_j \|\mathbf{r}_{l-1} - \mathbf{e}_{l,j}\|^2, \quad \mathbf{r}_l = \mathbf{r}_{l-1} - \mathbf{e}_{l,c_{l}},
\end{equation}
with $\mathbf{r}_0 = \mathbf{z}_i$. Here, $\mathbf{r}_{l-1}$ denotes the residual from the $(l{-}1)$-th layer, and $c_{l} \in \{1, \dots, N\}$ is the selected code index at layer $l$. This recursive process yields a semantic ID $\mathbf{c}_i = [c_{i,1}, \dots, c_{i,L}]$ for item $i$.
The reconstructed latent representation $\hat{\mathbf{z}}_i = \sum_{l=1}^{L} \mathbf{e}_{l,c_l}$ is then decoded back to the embedding space:
\begin{equation}
\hat{\mathbf{s}}_i = \mathrm{Decoder}(\hat{\mathbf{z}}_i), \quad \hat{\mathbf{z}}_i \in \mathbb{R}^{d_s}.
\end{equation}
The codebooks are optimized by minimizing the joint reconstruction and commitment loss:
\begin{equation}
    \mathcal{L}_{\text{RQ-VAE}} = \mathcal{L}_{\text{recon}} + \mathcal{L}_{\text{commit}},
\end{equation}
\begin{equation}
    \mathcal{L}_{\text{recon}} = \|\hat{\mathbf{s}}_i - \mathbf{s}_i\|^2,
\end{equation}
\begin{equation}
    \mathcal{L}_{\text{commit}} = \sum_{l=1}^{L} \|\mathrm{sg}(\mathbf{r}_{l-1}) - \mathbf{e}_{l,c_l}\|^2 + \mu \|\mathbf{r}_{l-1} - \mathrm{sg}(\mathbf{e}_{l,c_l})\|^2,
\end{equation}
where $\mathrm{sg}(\cdot)$ denotes the stop-gradient operation. 
The reconstruction loss $\mathcal{L}_{\text{recon}}$ ensures the reconstructed embedding retains the original semantics, and the commitment loss $\mathcal{L}_{\text{commit}}$ encourages proximity between 
the latent residuals and their assigned code embeddings.

\subsection{Topology-Aware Tokenization (TopoTok)}
\begin{figure*}[t]
  \centering
  \includegraphics[width=0.85\linewidth]{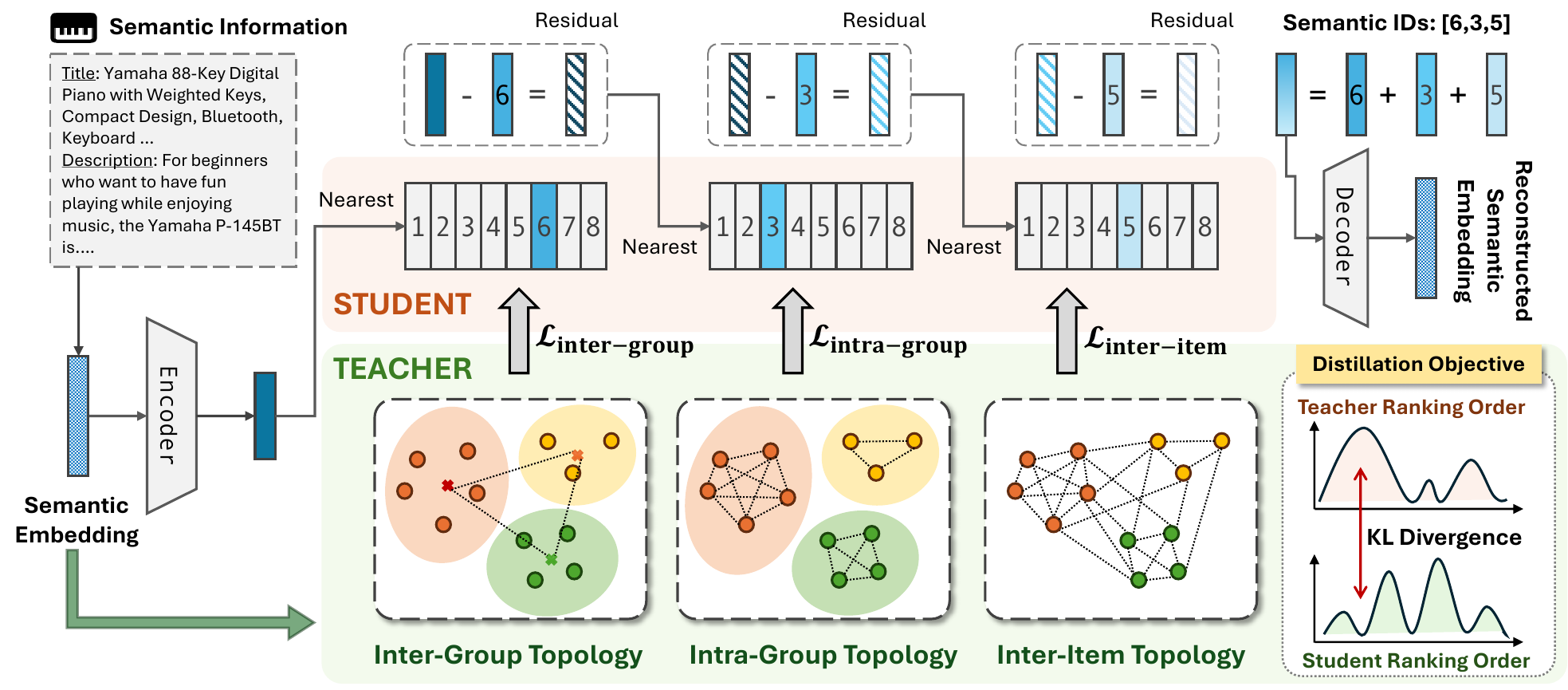}
  \vspace{-0.2cm}
  \caption{Illustration of TopoTok. TopoTok preserves topological relationships in semantic ID-based tokenization by imposing hierarchical, coarse-to-fine topology distillation across residual quantization layers. {A three-layer RQ-VAE is shown for illustration, while TopoTok generalizes to arbitrary depths.}}
  \Description{A comprehensive framework diagram of TopoTok consisting of two main parts: the semantic ID tokenization pipeline and the multi-granularity topology supervision mechanism. The top pipeline shows a Yamaha digital piano passing through a semantic information encoder to produce a continuous semantic embedding. This embedding undergoes a 3-layer recursive residual quantization via RQ-VAE to generate the discrete semantic ID sequence 6, 3, 5. The selected codebook embeddings are summed to create a reconstructed semantic embedding, which is processed by a decoder. The bottom part illustrates the distillation objectives between the teacher and student spaces across three hierarchical levels. First, Inter-Group Topology matches the ranking of group centroids at the first layer. Second, Intra-Group Topology aligns the local rankings of items within the same cluster at the second layer. Third, Inter-Item Topology enforces fine-grained global neighborhood ranking consistency at the final item level. All three levels optimize the student space by minimizing the KL divergence against the teacher ranking order.}
  \label{fig:pipeline}
  \vspace{-0.3cm}
\end{figure*}

{As shown in Figure~\ref{fig:pipeline}, we present TopoTok, which aligns topology supervision with the coarse-to-fine semantic progression of RQ-VAE by decomposing the distillation objective into three hierarchical levels: Inter-Group, Intra-Group, and Inter-Item. }

\subsubsection{\textbf{Topology Distillation Formulation}}

{In this section, we establish a mathematical framework for topology distillation. Formally, let $\{\mathbf{h}^{\text{t}}_i\}_{i=1}^{M}$ and $\{\mathbf{h}^{\text{s}}_i\}_{i=1}^{M}$ denote the representations of $M$ \emph{topological units} in the teacher and student spaces, respectively. A topological unit defines the basic entity for relational alignment.
To quantify the relational structure between units, we construct pairwise distance matrices $\mathbf{D}^{\text{t}}, \mathbf{D}^{\text{s}} \in \mathbb{R}^{M \times M}$, where each entry $d_{ij}$ measures the proximity between units $i$ and $j$ in the corresponding space. 
However, under quantization, information compression breaks the consistency of pairwise distances across spaces, making distance-level alignment ill-posed.
Therefore, we formulate topology distillation as a similarity ranking alignment problem, which focuses on preserving relative neighborhood structure. Specifically, we transform distances into similarity distributions by applying a row-wise softmax to the negated distance matrices:
\begin{equation}\label{eq:similarity}
\mathbf{P}^{\text{t}}[i, :] = \mathrm{softmax}(-\mathbf{D}^{\text{t}}[i, :]), \quad
\mathbf{P}^{\text{s}}[i, :] = \mathrm{softmax}(-\mathbf{D}^{\text{s}}[i, :]).
\end{equation}
Here, each row of $\mathbf{P}^{\text{t}}$ and $\mathbf{P}^{\text{s}}$ encodes the relative ranking of other units with respect to unit $i$ in the teacher and student spaces, respectively. By emphasizing neighbor ranking consistency rather than exact distance values, this formulation enables topology comparison across heterogeneous representation spaces.}

{Finally, topology distillation is achieved by minimizing the KL divergence between the teacher and student similarity distributions:
\begin{equation}\label{eq:loss}
\mathcal{L}_{\text{TD}} = \frac{1}{M} \sum_{i=1}^{M} \mathrm{KL}\left(\mathbf{P}^{\text{t}}[i, :] \,\|\, \mathbf{P}^{\text{s}}[i, :]\right),
\end{equation}
which encourages the student representations to preserve the neighborhood ranking structure defined in the teacher space, thereby maintaining topological structure during the tokenization.}

\subsubsection{\textbf{Multi-Granularity Topology Supervision}}

{This section details how TopoTok conducts topology supervision across multiple semantic granularities to achieve coarse-to-fine topology distillation that aligns with the hierarchical structure of RQ-VAE.}

\subsubsection*{\textbf{Inter-Group (IG)}}
{Inter-group level focuses on preserving coarse-grained topology at the group level, ensuring that high-level relationships among item groups in the continuous embedding space are distilled in the early-stage tokens learned by RQ-VAE.}

Given a residual layer $l$, each item $i$ is assigned to a semantic group based on its code index $c_{i,l}$ (Eq.~\ref{eq:residual_quant}). 
For each code $j \in \{1, \dots, N\}$ at layer $l$, we define the corresponding item group as:
\begin{equation}
\mathcal{G}_{l,j} = \{ i \mid c_{i,l} = j \},
\end{equation}
which encapsulates all items sharing high-level semantic features represented by the codebook embedding $\mathbf{e}_{l,j}$. 

At this level, each semantic group $\mathcal{G}_{l,j}$ serves as a topological unit. The teacher representation of $\mathcal{G}_{l,j}$ is defined as the group-level semantic centroid in the continuous embedding space:
\begin{equation}
\mathbf{h}^{\text{t}}_{j,\mathrm{IG}}
= \frac{1}{|\mathcal{G}_{l,j}|} \sum_{i \in \mathcal{G}_{l,j}} \mathbf{s}_i ,
\end{equation}
which summarizes the shared semantics of items within the group. 
The student representation of each semantic group is defined as the corresponding code embedding $\mathbf{h}^{\text{s}}_{j,\mathrm{IG}} = \mathbf{e}_{l,j}$, serving as the group-level prototype in the tokenized space.

To model the topological structure between item groups, we construct the pairwise inter-group distance matrices $\mathbf{D}^{\text{t}}_\mathrm{IG}$ and $\mathbf{D}^{\text{s}}_\mathrm{IG}$ in the teacher and student spaces, where each entry is defined as:
\begin{equation}
d^{\text{t}}_{jk} = \|\mathbf{h}^{\text{t}}_{j,\mathrm{IG}} - \mathbf{h}^{\text{t}}_{k,\mathrm{IG}}\|^2, \quad
d^{\text{s}}_{jk} = \|\mathbf{h}^{\text{s}}_{j,\mathrm{IG}} - \mathbf{h}^{\text{s}}_{k,\mathrm{IG}}\|^2,
\end{equation}
which encode the global relative arrangement of semantic groups.

By instantiating the topology distillation objective (Eqs.~\ref{eq:similarity} and~\ref{eq:loss}) on inter-group distance matrices, this level enforces consistency in neighborhood rankings among coarse-grained item groups across the embedding and tokenized spaces, thereby anchoring global group-level structure and preventing early-stage topological errors from propagating through the quantization hierarchy.

\subsubsection*{\textbf{Intra-Group (IaG)}}
{While inter-group distillation stabilizes coarse semantic organization, it leaves the internal structure of each group unconstrained. Intra-group distillation bridges this gap by refining local topology at an intermediate semantic granularity, where tokenized representations are enforced to capture fine-grained relationships among items within each group.}

At this level, each item $i$ within semantic groups serves as a topological unit. At the residual layer $l$ applied intra-group distillation, the teacher representation of item $i$ is defined as the original item embedding, and the student representation is the cumulative reconstructed representation up to the quantization depth $l$:
\begin{equation}
\mathbf{h}^{\text{t}}_{i,\mathrm{IaG}} = \mathbf{s}_i, \quad
\mathbf{h}^{\text{s}}_{i,\mathrm{IaG}} = \sum_{m=1}^{l} \mathbf{e}_{m,c_{i,m}} .
\end{equation}

To focus topology supervision on local structure, we restrict distance computation to item pairs belonging to the same semantic group determined at the preceding layer $l-1$  (Eq.~\ref{eq:residual_quant}). The teacher and student intra-group distance matrices are defined as:
\begin{align}
d^{\text{t}}_{ij} &=
\begin{cases}
\| \mathbf{h}^{\text{t}}_{i,\mathrm{IaG}} - \mathbf{h}^{\text{t}}_{j,\mathrm{IaG}} \|^2, 
& \text{if } c_{i,l-1} = c_{j,l-1} \text{ and } i \neq j, \\
\infty, & \text{otherwise},
\end{cases} \\
d^{\text{s}}_{ij} &=
\begin{cases}
\| \mathbf{h}^{\text{s}}_{i,\mathrm{IaG}} - \mathbf{h}^{\text{s}}_{j,\mathrm{IaG}} \|^2, 
& \text{if } c_{i,l-1} = c_{j,l-1} \text{ and } i \neq j, \\
\infty, & \text{otherwise},
\end{cases}
\end{align}
Here, distances for item pairs from different groups are set to $\infty$, ensuring that the topology supervision in Eq.~\ref{eq:similarity} focuses exclusively on intra-group neighborhood structure without affecting global group-level structure.
By integrating intra-group distances into the foundational template (Eqs.~\ref{eq:similarity} and~\ref{eq:loss}), intra-group distillation reinforces neighborhood rankings within each semantic group. This level is crucial for recovering fine-grained local structure that is often attenuated during item tokenization, thereby enhancing the model’s ability to discriminate between closely related items.

\subsubsection*{\textbf{Inter-Item (II)}}

{At deeper stages of RQ-VAE, we introduce inter-item distillation to preserve the global neighborhood structure at the fine-grained item level.}
Each item $i$ serves as a topological unit. The teacher representation is the original semantic embedding, while the student representation corresponds to the reconstructed representation aggregated up to the given quantization depth $l$:
\begin{equation}
\mathbf{h}^{\text{t}}_{i,\mathrm{II}} = \mathbf{s}_i, \quad
\mathbf{h}^{\text{s}}_{i,\mathrm{II}} = \sum_{m=1}^{l} \mathbf{e}_{m,c_{i,m}} ,
\end{equation}
where $l$ denotes the layer depth at which item-level topology supervision is applied.
To capture the global item-level topological structure, we construct pairwise inter-item distance matrices:
\begin{equation}
d^{\text{t}}_{ij} = \| \mathbf{h}^{\text{t}}_{i,\mathrm{II}} - \mathbf{h}^{\text{t}}_{j,\mathrm{II}} \|^2, \quad
d^{\text{s}}_{ij} = \| \mathbf{h}^{\text{s}}_{i,\mathrm{II}} - \mathbf{h}^{\text{s}}_{j,\mathrm{II}} \|^2 .
\end{equation}
These matrices encode the comprehensive relational structure among items in the embedding and tokenized spaces.

By substituting inter-item distance matrices into Eqs.~\ref{eq:similarity} and~\ref{eq:loss}, inter-item distillation enforces global neighborhood ranking consistency at the finest semantic granularity. 

\subsubsection{\textbf{Hierarchical Layer Deployment}}\label{sec:layer_assign}
{This section details how distillation at different granularities is assigned across the quantization layers to align with the hierarchical structure of RQ-VAE.}

Let $l \in \{1, \dots, L\}$ denote the index of a residual quantization layer in an $L$-layer RQ-VAE. 
We map the three topology distillation levels according to the semantic granularity captured at each layer:
\begin{itemize}[leftmargin=*]
\item \textbf{Inter-group distillation} is anchored at the first residual layer ($l=1$), where items are encoded into coarse semantic clusters. This distillation level establishes a global structure backbone by aligning neighborhood relations among high-level item groups.

\item \textbf{Intra-group distillation} is applied at the second layer ($l=2$), where representations begin to differentiate within established groups. 
Topology supervision at this stage refines local manifold structure while preserving group-level boundaries.

\item \textbf{Inter-item distillation} is applied at deeper layers ($2 < l \leq L$), where residual codes encode fine-grained semantic variations. At this stage, topology supervision operates directly at the item level to preserve global point-to-point neighborhood consistency, completing the coarse-to-fine alignment process.

\end{itemize}

In special configurations such as a two-layer RQ-VAE ($L=2$), we propose an inter-group and inter-item combination. This configuration preserves both global structure and item-level discriminability while respecting the hierarchical nature of residual quantization.

Beyond this fixed assignment, we further recommend using the codebook utilization rate as a practical indicator of semantic granularity to flexibly determine the optimal distillation level for each layer.
Low-utilization layers, where items concentrate on a small subset of codes, encode coarse semantics and are well-suited for inter-group distillation. Moderately utilized layers capture intermediate semantics and benefit from intra-group distillation. Highly utilized layers, where most or all codes are activated, encode fine residual structure and are best supervised by inter-item distillation. Typically, utilization rates in RQ-VAE tokenizers increase with depth, supporting our progressive deployment.

Overall, this stage-based deployment makes TopoTok architecture-agnostic and readily applicable to RQ-VAE frameworks of arbitrary depth. By aligning distillation granularity with the quantization hierarchy, TopoTok provides a general solution for generative recommendation models built on learnable hierarchical tokenization, achieving structural flexibility without architectural changes.

\subsubsection{\textbf{Training Objective}}

The proposed TopoTok is integrated into RQ-VAE backbone to guide the training of a topology-aware semantic ID tokenizer. The overall training objective is defined as:
\begin{equation}
\mathcal{L}_{\text{total}} = \mathcal{L}_{\text{RQ-VAE}} + \alpha \cdot \mathcal{L}_{\text{TopoTok}},
\end{equation}
where $\alpha$ is the topology distillation weight that controls the strength of topology supervision.
The TopoTok distillation loss is given by:
\begin{equation}
\mathcal{L}_{\text{TopoTok}} = \mathcal{L}_{\text{inter-group}} +  \mathcal{L}_{\text{intra-group}} + \mathcal{L}_{\text{inter-item}},
\end{equation}
This joint training objective enables RQ-VAE to preserve both item semantic information and the hierarchical topology information among items, resulting in higher-quality tokenization that better supports generative recommendation. 
For training complexity, TopoTok relies solely on batch-local, highly parallelizable computations, incurring a modest training overhead.
{Importantly, TopoTok introduces \textbf{no additional computation at inference time}, as the distillation objectives are only applied during training and do not alter the tokenization or recommendation pipeline at serving time. This property is particularly desirable in recommender systems, where inference latency is critical for real-time deployment.}

\begin{table*}[h]
\centering
\caption{Performance comparison across three datasets. The best and second-best results within each comparison group are highlighted in bold and underlined font, respectively. 
Superscript $^*$ indicates statistical significance at $p < 0.05$.
}
\vspace{-0.2cm}
\label{tab:main}
\resizebox{\textwidth}{!}{
\begin{tabular}{c|cccc|cccc|cccc}
\toprule
\multirow{2}{*}{\textbf{Model}} &
\multicolumn{4}{c|}{\textbf{Scientific}} &
\multicolumn{4}{c|}{\textbf{Instrument}} &
\multicolumn{4}{c}{\textbf{Game}} \\
& \textbf{R@5} & \textbf{R@10} & \textbf{N@5} & \textbf{N@10} & \textbf{R@5} & \textbf{R@10} & \textbf{N@5} & \textbf{N@10} & \textbf{R@5} & \textbf{R@10} & \textbf{N@5} &\textbf{ N@10} \\
\midrule
Caser            & 0.0172 & 0.0281 & 0.0107 & 0.0142 & 0.0242 & 0.0392 & 0.0154 & 0.0202 & 0.0346 & 0.0567 & 0.0221 & 0.0291 \\
GRU4Rec          & 0.0221 & 0.0353 & 0.0144 & 0.0186 & 0.0345 & 0.0537 & 0.0220 & 0.0281 & 0.0522 & 0.0831 & 0.0337 & 0.0436 \\
SASRec            & 0.0256 & 0.0406 & 0.0147 & 0.0195 & 0.0341 & 0.0530 & 0.0217 & 0.0277 & 0.0517 & 0.0821 & 0.0329 & 0.0426 \\
BERT4Rec          & 0.0180 & 0.0300 & 0.0113 & 0.0151 & 0.0305 & 0.0483 & 0.0196 & 0.0253 & 0.0453 & 0.0716 & 0.0294 & 0.0378 \\
FDSA              & 0.0261 & 0.0391 & 0.0174 & 0.0216 & 0.0364 & 0.0557 & 0.0233 & 0.0295 & 0.0548 & 0.0857 & 0.0353 & 0.0453 \\
S$^3$Rec          & 0.0253 & 0.0410 & 0.0172 & 0.0218 & 0.0340 & 0.0538 & 0.0218 & 0.0282 & 0.0533 & 0.0823 & 0.0351 & 0.0444 \\
P5-SID            & 0.0155 & 0.0234 & 0.0103 & 0.0129 & 0.0319 & 0.0438 & 0.0237 & 0.0275 & 0.0480 & 0.0693 & 0.0333 & 0.0401 \\
P5-CID            & 0.0192 & 0.0300 & 0.0123 & 0.0158 & 0.0352 & 0.0507 & 0.0234 & 0.0285 & 0.0497 & 0.0748 & 0.0343 & 0.0424 \\
\midrule
TIGER              & 0.0275 & 0.0431 & \underline{0.0181} & \underline{0.0231} & 0.0368 & 0.0574 & 0.0242 & 0.0308 & 0.0570 & 0.0895 & 0.0370 & 0.0471 \\
LETTER         & \underline{0.0276} & \underline{0.0433} & 0.0179 & 0.0230 & \underline{0.0372} & \underline{0.0581} & \underline{0.0243} & \underline{0.0310} & \underline{0.0576} & \underline{0.0901} & 0.0373 & 0.0475 \\
CoST           & 0.0270 & 0.0426 & 0.0180 & 0.0229 & 0.0366 & 0.0570 & 0.0242 & 0.0306 & 0.0569 & 0.0897 & \underline{0.0379} & \underline{0.0472} \\
\textbf{TIGER-TopoTok } & \textbf{0.0302}* & \textbf{0.0465}* & \textbf{0.0196}* & \textbf{0.0248}* & \textbf{0.0402}* & \textbf{0.0613}* & \textbf{0.0263}* & \textbf{0.0331}* & \textbf{0.0608}* & \textbf{0.0939}* & \textbf{0.0403}* & \textbf{0.0509}* \\
\midrule
\textit{Improv.} & +9.42\% & +7.39\% & +8.29\% & +7.36\% & +8.06\% & +5.51\% & +8.23\% & +6.77\% & +5.56\% & +4.22\% & +6.33\% & +7.84\% \\
\midrule
ETEGRec & \underline{0.0294} & \underline{0.0455} & \underline{0.0190} & \underline{0.0241} 
        & \underline{0.0402} & \underline{0.0624} & \underline{0.0260} & \underline{0.0331} 
        & \underline{0.0616} & \underline{0.0947} & \underline{0.0400} & \underline{0.0507} \\
\textbf{ETEGRec-TopoTok} & \textbf{0.0307*} & \textbf{0.0481*} & \textbf{0.0201*} & \textbf{0.0255*} & \textbf{0.0426*} & \textbf{0.0657*} & \textbf{0.0273*} & \textbf{0.0349*} & \textbf{0.0635*} & \textbf{0.0975*} & \textbf{0.0414*} & \textbf{0.0525*} \\
\midrule
\textit{Improv.} & +4.42\% & +5.71\% & +5.79\% & +5.81\% & +5.97\% & +5.29\% & +5.00\% & +5.44\% & +3.08\% & +2.96\% & +3.50\% & +3.55\%\\
\bottomrule
\end{tabular}
}
\end{table*}

\section{EXPERIMENTS}

\subsection{Experimental Setting}

\subsubsection{Dataset}
We evaluate our method following the standard protocol used in prior work~\cite{tiger, letter}. Experiments are conducted on three subsets of the latest Amazon Review dataset~\cite{hou2024bridging}: \emph{Industrial Scientific}, \emph{Musical Instruments}, and \emph{Video Games}. We apply a 5-core filtering procedure, removing users and items with fewer than five interactions. User interaction sequences are then constructed in chronological order, with a maximum sequence length of 20. 

\subsubsection{Baseline Models}
We compare TopoTok with comprehensive baselines from three categories: 
(1) Traditional sequential recommendation methods: \textbf{Caser}~\cite{caser}, \textbf{GRU4Rec}~\cite{gru4rec}, \textbf{SASRec}~\cite{sasrec}, \textbf{BERT4Rec}~\cite{bert4rec}, \textbf{FDSA}~\cite{fdsa}, and \textbf{S$^3$Rec}~\cite{s3rec}. 
(2) Generative recommendation methods: \textbf{P5-CID}~\cite{index}, \textbf{P5-SID}~\cite{index}, \textbf{TIGER}~\cite{tiger}, and \textbf{ETEGRec}~\cite{etegrec}. 
(3) Semantic ID tokenization enhancement methods: \textbf{LETTER}~\cite{letter} and \textbf{CoST}~\cite{cost} (both implemented on TIGER).

\subsubsection{Evaluation Protocol and Implementation Details}
We evaluate all models using top-K Recall (R@K) and NDCG (N@K) with \( K = \{5, 10\} \). Following standard practice~\cite{tiger}, we adopt the \textit{leave-one-out} strategy: for each user, the last interaction is used for testing, the second-last for validation, and the rest for training. We conduct a full-ranking evaluation over the entire candidate item set without sampling. 
We adopt TIGER~\cite{tiger} and ETEGRec~\cite{etegrec} as backbone generative recommenders, using Sentence-T5~\cite{ni2022sentence} and SASRec~\cite{sasrec} to obtain item embeddings, respectively. 
For item tokenization, we use RQ-VAE with three codebook layers (each with 256 codes of dimension 128, except for Section~\ref{sec:exp_depth}). TopoTok is trained for 10k epochs using AdamW~\cite{adamw} (lr=1e-3, batch size=2048), with the topology weight $\alpha \in \{0.01, 0.1, 0.3, 0.5, 1\}$ selected on validation. 
Following TIGER~\cite{tiger}, we append an additional token to ensure semantic ID uniqueness. We use T5 as our recommender and follow the original training protocols~\cite{tiger,etegrec}. 
All experiments are conducted on a single NVIDIA Tesla A40 GPU.
Results are reported using the model with the best validation NDCG@10. Statistical significance is assessed via a paired t-test over five independent runs. For the main results in Table~\ref{tab:main}, we report performance using seed 2025 for reproducibility. Unless otherwise specified, experiments are conducted on the TIGER backbone.

\vspace{-0.1cm}
\subsection{Overall Performance}\label{sec:overall}

{We evaluate TopoTok under two representative generative recommendation paradigms: TIGER~\cite{tiger}, which trains the tokenizer and recommender sequentially, and ETEGRec~\cite{etegrec}, which adopts end-to-end training. Following their original designs, tokenization enhancement methods such as LETTER~\cite{letter} and CoST~\cite{cost} are implemented under the TIGER backbone and included in the TIGER-based comparison. The overall results are in Table~\ref{tab:main}.}

{\scalebox{1.5}{\textbullet}\ \textbf{TopoTok consistently improves generative recommendation across both backbones.}  
Under the TIGER backbone, TIGER-TopoTok achieves the best performance across all datasets and metrics, delivering statistically significant gains over TIGER and prior tokenization enhancement methods such as LETTER and CoST (e.g., up to $+9.42\%$ on Scientific). When integrated into the end-to-end ETEGRec framework, ETEGRec-TopoTok further establishes new state-of-the-art results. These results demonstrate that TopoTok is robust in both two-stage and end-to-end settings.}

{\scalebox{1.5}{\textbullet}\ \textbf{End-to-end training with TopoTok unlocks the full potential of RQ-VAE tokenization.}  
ETEGRec-TopoTok consistently surpasses TIGER-TopoTok across all datasets, demonstrating the benefit of jointly optimizing item tokenization and generation. This advantage stems from the parameterized nature of RQ-VAE, which allows gradients from the generative objective to propagate back to the tokenizer. By providing topology-aware supervision throughout the quantization hierarchy, TopoTok stabilizes end-to-end optimization and enables RQ-VAE to preserve relational structure while adapting token representations to downstream generation.}

{\scalebox{1.5}{\textbullet}\ \textbf{Topology supervision must respect the hierarchical structure of residual quantization.}  
Within the TIGER-based group, although LETTER and CoST provide improvements, their gains are limited by a lack of hierarchical awareness. Specifically, LETTER does not incorporate explicit topology supervision, while CoST applies monolithic contrastive supervision only at the final output, overlooking the layer-wise semantic progression of RQ-VAE. 
We further observe that CoST exhibits inconsistent performance across datasets, which we attribute to its limited ability to preserve topology under hierarchical semantics, as supported by the analysis in Section~\ref{sec:vis_study}. Moreover, applying contrastive supervision only at the final layer may interfere with reconstruction learning, leading to degraded performance in Table~\ref{tab:main}.
In contrast, TopoTok explicitly aligns topology supervision with the coarse-to-fine structure of RQ-VAE, enabling each codebook layer to preserve topology at an appropriate semantic scale.}

\subsection{Ablation Study}

\subsubsection{Multi-Granularity Topology Distillation}
To evaluate the effectiveness of our hierarchical topology distillation, we conduct ablation studies by incrementally adding Inter-Group (IG), Intra-Group (IaG), and Inter-Item (II) Distillation on the TIGER backbone. We also report the top-20 neighborhood overlap across the three codebook layers, which is computed as the average ratio of shared top-20 nearest neighbors between the original semantic embeddings and the reconstructed representations at each layer. Higher overlap indicates better topology preservation during tokenization. 
From the results in Table~\ref{tab:ablation_combined}, we derive the following conclusions:

\scalebox{1.5}{\textbullet}\ \textbf{The three levels of topology distillation are complementary and mutually reinforcing.} For example, introducing IG alone leads to notable gains in overlap across all layers, suggesting that coarse-grained topology supervision at the early layer induces structural adjustments that mitigate the accumulation of topology distortion in deeper layers. Conversely, adding supervision at deeper layers also leads to improvements in earlier layers. 

\scalebox{1.5}{\textbullet}\ \textbf{TopoTok alleviates topology distortion.} The full TopoTok achieves the highest neighborhood overlaps across all datasets, confirming its ability to preserve multi-level topological structure. The corresponding improvements in recommendation performance further support the conclusion that reducing topology distortion enhances token quality and benefits downstream generation.

\begin{table}[t]
\centering
\large
\caption{Ablation study of TopoTok components. \textbf{IG}, \textbf{IaG}, and \textbf{II} denote Inter-Group, Intra-Group, and Inter-Item distillation, respectively. l indicates the codebook layer.}
\vspace{-0.3cm}
\label{tab:ablation_combined}
\resizebox{1.02\linewidth}{!}{
\begin{tabular}{c|ccc|cccc|ccc}
\toprule
 & \multicolumn{3}{c|}{\textbf{Variants}} & \multicolumn{4}{c|}{\textbf{Metrics}} & \multicolumn{3}{c}{\textbf{Top-20 Overlap}} \\
\cmidrule(lr){2-4} \cmidrule(lr){5-8} \cmidrule(lr){9-11}
& \textbf{IG} & \textbf{IaG} & \textbf{II} & \textbf{R@5} & \textbf{R@10} & \textbf{N@5} & \textbf{N@10} & \textbf{l1} & \textbf{l2} & \textbf{l3} \\
\midrule
\multirow{6}{*}{\rotatebox{90}{\textbf{Scientific}}} 
& - & - & - & 0.0275 & 0.0431 & 0.0181 & 0.0231 & 62.90\% & 30.29\% & 27.55\% \\
& \cmark & - & - & 0.0288 & 0.0450 & 0.0187 & 0.0239 & 72.47\% & 35.99\% & 31.43\% \\
& - & \cmark & - & 0.0280 & 0.0439 & 0.0183 & 0.0234 & 66.12\% & 32.88\% & 32.05\% \\
& - & - & \cmark & 0.0285 & 0.0443 & 0.0189 & 0.0240 & 66.29\% & 31.54\% & 32.94\% \\
& \cmark & \cmark & - & 0.0293 & 0.0461 & 0.0193 & 0.0243 & 73.46\% & 39.05\% & 32.56\% \\
& \cmark & \cmark & \cmark & \textbf{0.0302} & \textbf{0.0465} & \textbf{0.0196} & \textbf{0.0248} & \textbf{73.70\%} & \textbf{41.26\%} & \textbf{36.75\%} \\
\midrule
\multirow{6}{*}{\rotatebox{90}{\textbf{Instrument}}} 
& - & - & - & 0.0368 & 0.0574 & 0.0242 & 0.0308 & 63.06\% & 27.62\% & 26.49\% \\
& \cmark & - & - & 0.0386 & 0.0590 & 0.0259 & 0.0324 & 65.69\% & 31.13\% & 27.22\% \\
& - & \cmark & - & 0.0391 & 0.0606 & 0.0259 & 0.0328 & 64.27\% & 29.86\% & 31.71\% \\
& - & - & \cmark & 0.0382 & 0.0586 & 0.0252 & 0.0318 & 63.06\% & 27.62\% & 30.09\% \\
& \cmark & \cmark & - & 0.0397 & 0.0609 & \textbf{0.0263} & 0.0330 & 72.07\% & 33.12\% & 28.78\% \\
& \cmark & \cmark & \cmark & \textbf{0.0402} & \textbf{0.0613} & \textbf{0.0263} & \textbf{0.0331} & \textbf{74.05\%} & \textbf{34.60\%} & \textbf{29.20\%} \\
\midrule
\multirow{6}{*}{\rotatebox{90}{\textbf{Game}}} 
& - & - & - & 0.0570 & 0.0895 & 0.0370 & 0.0471 & 55.08\% & 27.97\% & 27.57\% \\
& \cmark & - & - & 0.0587 & 0.0912 & 0.0387 & 0.0491 & 57.12\% & 30.49\% & 29.92\% \\
& - & \cmark & - & 0.0588 & 0.0914 & 0.0388 & 0.0492 & 58.27\% & 32.45\% & 27.57\% \\
& - & - & \cmark & 0.0582 & 0.0908 & 0.0381 & 0.0485 & 55.08\% & 27.75\% & 29.29\% \\
& \cmark & \cmark & - & 0.0595 & 0.0931 & 0.0391 & 0.0500 & 61.61\% & 31.90\% & 30.74\% \\
& \cmark & \cmark & \cmark & \textbf{0.0608} & \textbf{0.0939} & \textbf{0.0403} & \textbf{0.0509} & \textbf{63.30\%} & \textbf{35.67\%} & \textbf{33.36\%} \\
\bottomrule
\end{tabular}
}
\end{table}

\begin{table}[t]
\centering
\caption{Robustness of TopoTok across different RQ-VAE layers (\textbf{L}). \textbf{Bold} indicates the better result for each configuration.}
\vspace{-0.3cm}
\label{tab:ablation_layers}
\resizebox{0.48\textwidth}{!}{
\begin{tabular}{c|c|c|cc|c}
\toprule
\textbf{Dataset} & \textbf{Layers (L)} & \textbf{Model} & \textbf{Recall@10} & \textbf{NDCG@10} & \textbf{Improv.} \\ 
\midrule
\multirow{6}{*}{\textbf{Scientific}} 
& \multirow{2}{*}{\textbf{L=2}} & Base & 0.0400 & 0.0213 & \multirow{2}{*}{+11.25\%} \\
& & \textbf{+TopoTok} & \textbf{0.0445} & \textbf{0.0237} & \\
\cmidrule{2-6}
& \multirow{2}{*}{\textbf{L=3}} & Base & 0.0431 & 0.0231 & \multirow{2}{*}{+7.89\%} \\
& & \textbf{+TopoTok} & \textbf{0.0465} & \textbf{0.0248} & \\
\cmidrule{2-6}
& \multirow{2}{*}{\textbf{L=4}} & Base & 0.0457 & 0.0246 & \multirow{2}{*}{+2.41\%} \\
& & \textbf{+TopoTok} & \textbf{0.0468} & \textbf{0.0249} & \\
\midrule
\multirow{6}{*}{\textbf{Instrument}} 
& \multirow{2}{*}{\textbf{L=2}} & Base & 0.0569 & 0.0303 & \multirow{2}{*}{+4.57\%} \\
& & \textbf{+TopoTok} & \textbf{0.0595} & \textbf{0.0322} & \\
\cmidrule{2-6}
& \multirow{2}{*}{\textbf{L=3}} & Base & 0.0574 & 0.0308 & \multirow{2}{*}{+6.79\%} \\
& & \textbf{+TopoTok} & \textbf{0.0613} & \textbf{0.0331} & \\
\cmidrule{2-6}
& \multirow{2}{*}{\textbf{L=4}} & Base & 0.0601 & 0.0329 & \multirow{2}{*}{+2.66\%} \\
& & \textbf{+TopoTok} & \textbf{0.0617} & \textbf{0.0336} & \\
\bottomrule
\end{tabular}
}
\vspace{-0.5cm}
\end{table}

\subsubsection{Quantization Depth}\label{sec:exp_depth}

{We conduct an ablation study to examine the robustness of TopoTok under different RQ-VAE quantization depths. Experiments are performed on TIGER backbone, with the number of residual layers set to the commonly adopted configurations $L \in \{2,3,4\}$. The layer-wise deployment of topology distillation follows the strategy described in Section~\ref{sec:layer_assign}. Performance improvements are computed based on Recall@10.}

{As shown in Table~\ref{tab:ablation_layers}, TopoTok consistently improves recommendation performance across all tested quantization depths and datasets, demonstrating stability across layer configurations. The relative gains are most pronounced when $L=2$, where the shallow RQ-VAE hierarchy leads to greater information loss during residual quantization. In this setting, topology supervision provided by TopoTok effectively compensates for the limited representational capacity. As the number of residual layers increases ($L=3,4$), the tokenizer captures progressively finer semantic structure, and TopoTok continues to deliver stable improvements, confirming its compatibility with deeper hierarchical tokenization.
These results show that by aligning topology supervision with the semantic roles of residual layers, TopoTok remains effective across varying RQ-VAE depths, highlighting its generality and practical applicability.}

\subsection{Hyperparameter Analysis}

\begin{figure}[t]
  \setlength{\belowcaptionskip}{-0.5em}
  \centering
  \includegraphics[width=1.05\linewidth]{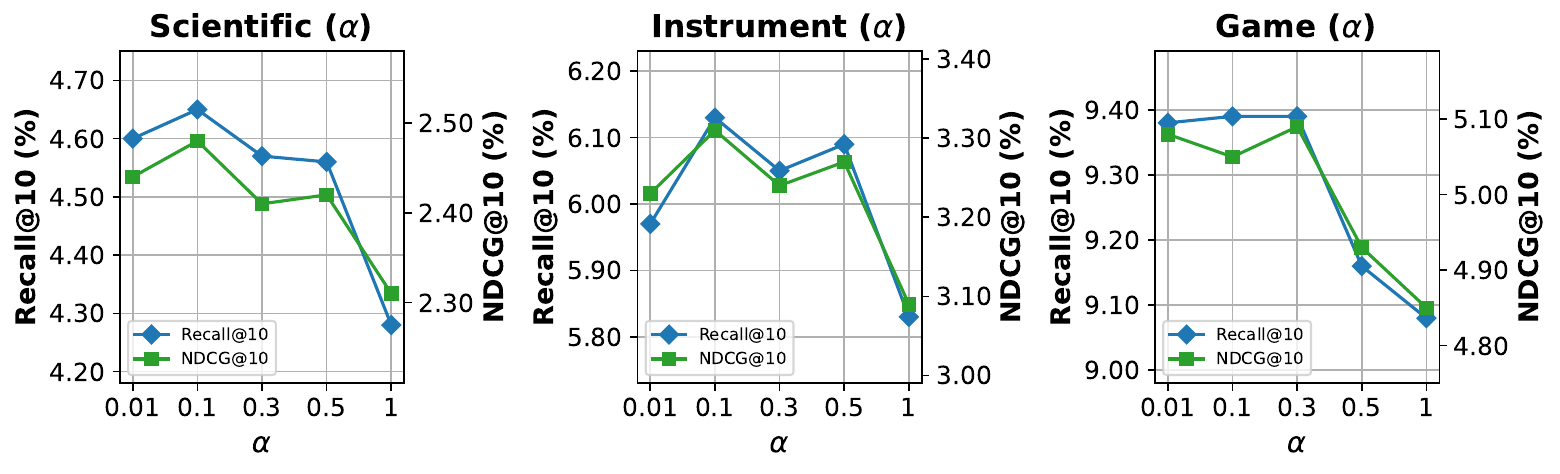}
  \vspace{-0.8cm}
  \caption{Performance across different $\alpha$ on three datasets.}
  \Description{Three line charts illustrating the performance of TopoTok across different topology distillation weights, alpha, valued at 0.01, 0.1, 0.3, 0.5, and 1. The left chart shows performance on the Scientific dataset, where Recall@10 and NDCG@10 peak at alpha equals 0.1 and then decline. The middle chart displays the Instrument dataset, which follows an identical trend, peaking sharply at alpha equals 0.1. The right chart represents the Game dataset, where performance peaks slightly later at alpha equals 0.3 before decreasing as the regularization weight increases further.}
  \label{fig:hyp1}
  \vspace{-0.1cm}
\end{figure}

We conduct controlled experiments on three datasets to examine the effect of the topology distillation weight $\alpha$, with results shown in Figure~\ref{fig:hyp1}. As $\alpha$ increases from small values, both Recall@10 and NDCG@10 consistently improve across datasets, indicating that introducing topology supervision effectively guides the tokenizer to preserve relational structure. Performance peaks at moderate values of $\alpha$, after which further increasing $\alpha$ leads to a clear decline. This trend suggests that overly strong topology constraints introduce excessive regularization, which hampers semantic reconstruction during vector quantization. The optimal $\alpha$ is dataset-dependent: $\alpha=0.1$ yields the best performance on the Scientific and Instrument datasets, while $\alpha=0.3$ is optimal for the Game dataset. Overall, these results highlight the importance of balancing topology supervision with reconstruction objectives to achieve optimal tokenization quality.

\subsection{Visualization Case Study}
\label{sec:vis_study}

While each distillation objective of TopoTok improves topology preservation (as shown in Table~\ref{tab:ablation_combined}), we further qualitatively assess the impact of topology-aware distillation comparing TIGER, CoST, and TIGER-TopoTok.
Specifically, we conduct a visualization case study comparing TIGER, CoST, and TIGER-TopoTok. The goal is to examine how well the neighborhood structure in the semantic space is preserved across quantization layers. 

\scalebox{1.5}{\textbullet}\ \textit{Experimental Setup.}
We randomly select a query item from the Instrument dataset and retrieve its top-20 nearest neighbors in the semantic space and the tokenized space at three quantization layers, forming a subset of sampled items for analysis. 
The resulting rank pairs are visualized in scatter plots, where the $x$-axis denotes the rank in the semantic space, and the $y$-axis denotes the rank in the tokenized space. The diagonal line represents perfect alignment between the two rankings, reflecting ideal topological preservation. 

The case study consists of two parts. First, we highlight the top-20 semantic neighbors of the query item in color, enabling a clear inspection of their rank consistency across layers. Second, to mitigate the randomness of the query item, we plot all rank pairs of other sampled items in gray, offering a broader view of general trends in topological preservation.

\begin{figure}[t]
  \setlength{\belowcaptionskip}{-0.5em}
  \centering
  \includegraphics[width=1.05\linewidth]{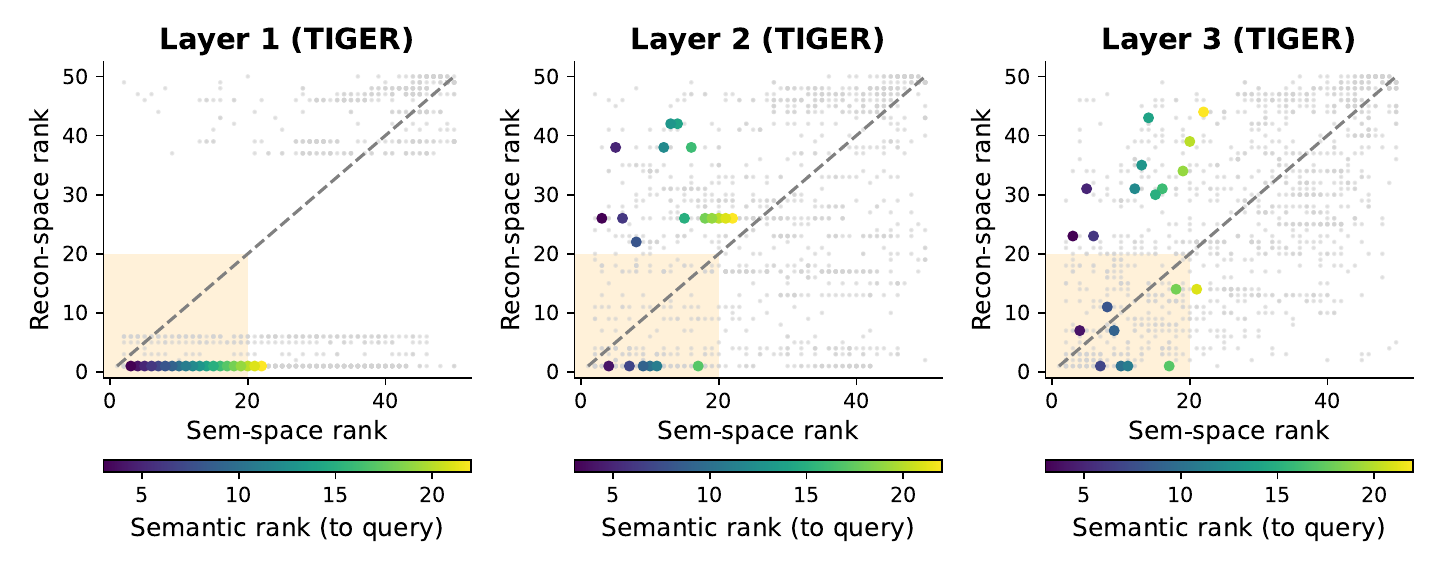}
  \vspace{-0.8cm}
  \caption{Rank comparison between semantic and reconstructed spaces at each layer in TIGER.}
  \Description{Three rank comparison scatter plots for the TIGER model across three quantization layers. The x-axis represents the semantic space rank, and the y-axis represents the reconstructed space rank. At Layer 1, colored points indicating top-20 semantic neighbors are well-aligned near the diagonal line. However, at Layer 2 and Layer 3, the colored points scatter widely away from the bottom-left region and deviate heavily from the diagonal, showing severe topology distortion as quantization depth increases.}
  \label{fig:scatter_tiger}
  \vspace{-0.2cm}
\end{figure}

\begin{figure}[t]
  \setlength{\belowcaptionskip}{-0.5em}
  \centering
  \includegraphics[width=1.05\linewidth]{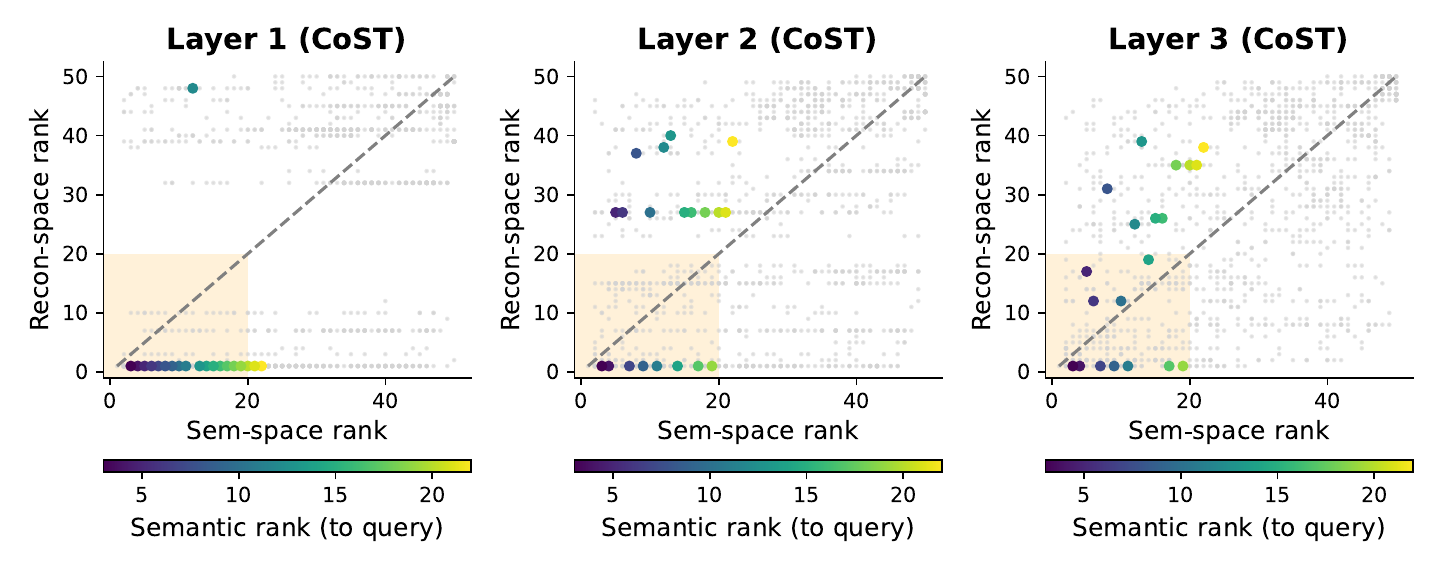}
  \vspace{-0.8cm}
  \caption{Rank comparison between semantic and reconstructed spaces at each layer in CoST.}
  \Description{Three rank comparison scatter plots for the CoST model across three quantization layers. Similar to TIGER, while some local neighborhood structure is retained at Layer 1 with colored points near the bottom-left corner, the rank consistency significantly deteriorates in Layer 2 and Layer 3. The points spread out loosely across the upper-right area, failing to maintain tight alignment along the diagonal line.}
  \label{fig:scatter_cost}
  \vspace{-0.2cm}
\end{figure}

\begin{figure}[t]
  \setlength{\belowcaptionskip}{-0.5em}
  \centering
  \includegraphics[width=1.05\linewidth]{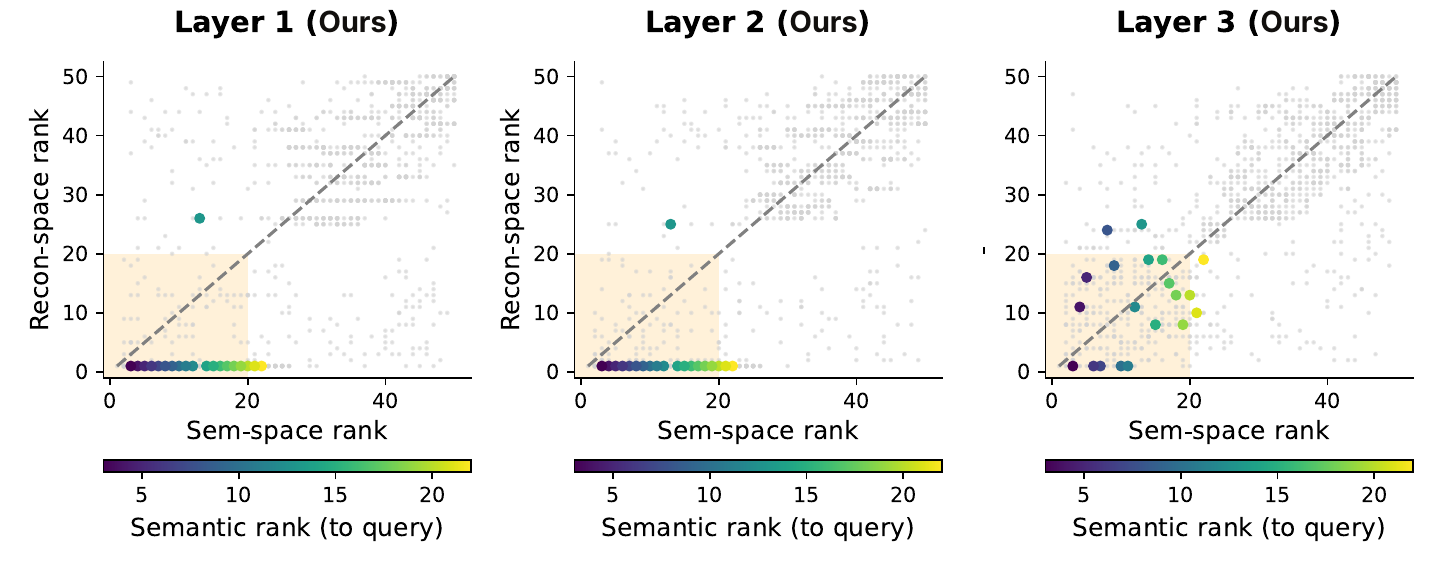}
  \vspace{-0.8cm}
  \caption{Rank comparison between semantic and reconstructed spaces at each layer in TopoTok.}
  \Description{Three rank comparison scatter plots for TopoTok across three quantization layers, demonstrating superior topological preservation. In Layer 1 and Layer 2, almost all colored points representing the top-20 semantic neighbors are tightly clustered at the absolute bottom-left corner of the plot, meaning they consistently maintain a top-1 rank in the reconstructed space. In Layer 3, the points remain highly concentrated and perfectly aligned along the ideal diagonal line, while the gray background points are also distributed significantly closer to the diagonal than the baselines.}
  \label{fig:scatter_TopoTok}
\end{figure}

\scalebox{1.5}{\textbullet}\ \textit{Interpretation Criteria.} 
A desirable result exhibits two key patterns. First, most colored points fall within or near the bottom-left highlighted region (rank $\leq 20$ in both spaces), indicating minimal distortion in the query item's top-20 neighborhood. Second, gray points concentrated along or near the diagonal, reflecting generally consistent relative rankings between the semantic and reconstructed spaces. Together, these patterns suggest that the semantic topology is well preserved during item tokenization.

\scalebox{1.5}{\textbullet}\ \textit{Results and Comparison.}
Figures~\ref{fig:scatter_tiger}, \ref{fig:scatter_cost}, and \ref{fig:scatter_TopoTok} present rank-scatter plots for TIGER, CoST, and TopoTok on the same query item, B09V188Y4X. The semantic IDs of the query item and its neighbors are reported in Tables~\ref{tab:semids_combined} for reference.

TIGER preserves rank alignment at layer 1, where most top-20 semantic neighbors share the same first semantic ID. However, this alignment degrades in deeper layers, with many neighbors falling outside the top-20 range and points increasingly deviating from the diagonal, indicating topology distortion.
CoST exhibits similar behavior: while some local structure is retained at layer 1, rank consistency deteriorates in layers 2 and 3. This behavior stems from applying a monolithic contrastive objective, without respecting the hierarchical semantics encoded across residual layers

In contrast, TopoTok consistently preserves topological structure across all layers. In layers 1 and 2, most of the top-20 semantic neighbors of the query item are ranked first, indicating that they share the same first two semantic IDs, 50 and 207. In layer 3, more neighbors remain within the top-20 reconstructed neighborhood, with points more tightly concentrated along the diagonal. Furthermore, the gray points are distributed closer to the diagonal, suggesting that the relational structure is more accurately preserved. These results demonstrate that explicitly aligning topology supervision with the coarse-to-fine quantization hierarchy enables TopoTok to achieve superior multi-level topology preservation.

\begin{table}[t]
\centering
\small
\caption{Semantic IDs of the query item and its top-20 semantic neighbors across TIGER, CoST, and TopoTok. Neighbors are sorted by ascending semantic-space rank (1–20).}
\vspace{-0.1cm}
\label{tab:semids_combined}
\begin{tabular}{c|l|l|l|l}
\toprule
\textbf{Rank} & \textbf{Item asin} & \textbf{IDs (TIGER)} & \textbf{IDs (CoST)} & \textbf{IDs (TopoTok)} \\
\midrule
Query & B09V188Y4X & [106, 56, 21]  & [116, 193, 145] & [50, 207, 136] \\
\midrule
1  & B093L2LHB9 & [106, 220, 0]  & [116, 193, 145] & [50, 207, 136] \\
2  & B089GTNJYQ & [106, 56, 0]   & [116, 193, 145] & [50, 207, 36]  \\
3  & B08372HW3L & [106, 51, 222] & [116, 65, 239]  & [50, 207, 76]  \\
4  & B083ZFH24H & [106, 220, 0]  & [116, 65, 31]   & [50, 207, 136] \\
5  & B085KW8K3F & [106, 56, 21]  & [116, 193, 145] & [50, 207, 136] \\
6  & B085XF93S7 & [106, 30, 21]  & [116, 233, 142] & [50, 207, 245] \\
7  & B085DM132N & [106, 56, 0]   & [116, 193, 145] & [50, 207, 17]  \\
8  & B08PK7CDKW & [106, 56, 21]  & [116, 65, 31]   & [50, 207, 136] \\
9  & B08H7Y1HQY & [106, 56, 21]  & [116, 193, 145] & [50, 207, 136] \\
10 & B082KZ3R2F & [106, 51, 222] & [252, 65, 37]   & [50, 207, 36]  \\
11 & B0B5QTG996 & [106, 192, 187]& [116, 57, 145]  & [44, 207, 50]  \\
12 & B07CZD8S8H & [106, 192, 7]  & [116, 193, 235] & [50, 207, 115] \\
13 & B07X5VT56K & [106, 220, 161]& [116, 65, 34]   & [50, 207, 109] \\
14 & B0823DMFG2 & [106, 51, 222] & [116, 65, 34]   & [50, 207, 115] \\
15 & B08P7CMR1Q & [106, 56, 21]  & [116, 193, 145] & [50, 207, 191] \\
16 & B0B1DJ7BB7 & [106, 220, 49] & [116, 65, 145]  & [50, 207, 24]  \\
17 & B07VT2YD88 & [106, 220, 210]& [116, 193, 145] & [50, 207, 109] \\
18 & B0B1DN5CSJ & [106, 220, 233]& [116, 65, 145]  & [50, 207, 24]  \\
19 & B0BKFZP9KR & [106, 220, 49] & [116, 65, 145]  & [50, 207, 190] \\
20 & B09S9SMDZK & [106, 220, 25] & [116, 73, 69]   & [50, 207, 115] \\
\bottomrule
\end{tabular}
\vspace{-0.4cm}
\end{table}

\section{CONCLUSION}
In this paper, we identify the underexplored topology distortion problem in existing semantic ID-based item tokenization for generative recommendation. To address this issue, we propose a novel \textbf{Topo}logy-Aware \textbf{Tok}enization framework (TopoTok), which decomposes topology supervision into three coarse-to-fine levels, aligning with the residual quantization hierarchy. Extensive experiments demonstrate that TopoTok significantly enhances topology preservation and outperforms state-of-the-art baselines.
TopoTok offers a general solution for preserving topological structure in item tokenization across various generative recommendation models.

\begin{acks} 
This research is supported in part by the National Science Foundation under Grant No. CNS-2427070,  IIS-2331069,  IIS-2202481, IIS-2130263, CNS-2131622. The views and conclusions contained in this document are those of the authors and should not be interpreted as representing the official policies, either expressed or implied, of the U.S. Government. The U.S. Government is authorized to reproduce and distribute reprints for Government purposes notwithstanding any copyright notation hereon.
\end{acks}

\bibliographystyle{ACM-Reference-Format}
\bibliography{reference}

\end{document}